\documentclass[11pt,fleqn]{article}

\usepackage{amsfonts}
\usepackage{amsmath}
\usepackage{amssymb}
\usepackage{mathtools}
\usepackage{mathrsfs}
\usepackage{enumerate}
\usepackage{bbm}
\usepackage{graphicx,epsfig,amsthm,color}
\usepackage{pstricks}
\usepackage{subfig}
\usepackage{graphicx}
\usepackage{units}
\usepackage{caption}
\usepackage{booktabs}
\usepackage{textcomp}
\usepackage{array}
\usepackage{cite}
\usepackage{cancel}

\date{\today}

\newcolumntype{z}[1]{>{\RaggedRight\hspace{0pt}}p{#1}}
\newcolumntype{w}[1]{>{\RaggedRight\hspace{0pt}}p{#1}}
\newcolumntype{v}[1]{>{\Centering\hspace{0pt}}p{#1}}

\def\be{\begin{equation}}
\def\ee{\end{equation}}
\def\bea{\begin{eqnarray}}
\def\eea{\end{eqnarray}}

\def\be{\begin{equation}}
\def\ee{\end{equation}}
\def\bea{\begin{eqnarray}}
\def\eea{\end{eqnarray}}

\def\erp2{{\rm e}^{2\rho}}
\def\erm2{{\rm e}^{-2\rho}}
\def\er4{{\rm e}^{4\rho}}

\def\be{\begin{equation}}
\def\ee{\end{equation}}
\def\bea{\begin{eqnarray}}
\def\eea{\end{eqnarray}}

\def\m0{m_{\nu_{0,i}}}

\def\T0{T_{\nu_0}}

\newcommand{\beqa}{\begin{eqnarray}}
\newcommand{\eeqa}{\end{eqnarray}}
\newcommand{\bpr}{\begin{problem}}
\newcommand{\epr}{\end{problem}}
\newcommand{\bcent}{\begin{center}}
\newcommand{\ecent}{\end{center}}
\newcommand{\bfig}{\begin{figure}}
\newcommand{\efig}{\end{figure}}
\newcommand{\bpc}{\begin{picture}}
\newcommand{\epc}{\end{picture}}

\renewcommand{\and}{A_{0}^{\nu ,D}(s)}

\newcommand{\bee}{\begin{equation}}

\def\beq{\begin{eqnarray}}
\def\eeq{\end{eqnarray}}

\newcommand{\bright}{\begin{flushright}}
\newcommand{\eright}{\end{flushright}}
\newcommand{\bminip}{\begin{minipage}}
\newcommand{\eminip}{\end{minipage}}

\DeclareCaptionLabelSeparator{mysep}{\hspace{3pt}:\hspace{3pt}}
\DeclareCaptionLabelFormat{mypiccap}{Fig.\hspace{3pt}{#2}}
\DeclareCaptionLabelFormat{mytabcap}{Tab.\hspace{3pt}{#2}}

\begin{document}

\date{}
\title{
\vskip 2cm {\bf\huge Optical theorem and effective action: \\
new proofs of old results in QFT}\\[0.8cm]}

\author{{\sc\normalsize
Sergio Luigi Cacciatori$^{1}$ and Andrea Zanzi$^{2}$\!
\!}\\[1cm]
{\normalsize $^{1}$ Department of Science and High Technology,}\\
{\normalsize Universit\`a dell'Insubria, Via Valleggio 11, I-22100, Como, Italy,}\\
{\normalsize and INFN, via Celoria 16, I-20133, Milano, Italy}\\
{\normalsize $^{2}$ La Gassa 17, CH-7017 Flims, Switzerland}\\
\\
{\normalsize Emails: sergio.cacciatori@uninsubria.it, andrea.zanzi@unife.it}\\[1cm]
}
\maketitle \thispagestyle{empty}

\begin{abstract}
{A new proof of the optical theorem at all orders is presented. Although the theorem is a well-known result in Quantum Field Theory, our proof is interesting because it is particularly simple. Indeed, the theorem is a direct consequence of the pole-ology formalism discussed in Weinberg's Cambridge books. We also discuss a new proof of the standard result concerning the effective action as generating functional of 1PI contributions.}
\end{abstract}
\clearpage

%%%%%%%%%%%%%%%%%%%%%%%%%%%%%%%%%%%%%%%%%%%%%%%%%%%%%%%%%%%%%%%%%%%%%%%%%%%%%%%%%%%%%%%%%%%%%%%%%%%%%%%%%%%%%%%%%%%%%
% Section INTRODUCTION
%%%%%%%%%%%%%%%%%%%%%%%%%%%%%%%%%%%%%%%%%%%%%%%%%%%%%%%%%%%%%%%%%%%%%%%%%%%%%%%%%%%%%%%%%%%%%%%%%%%%%%%%%%%%%%%%%%%%%
\setcounter{equation}{0}
\section{Introduction}

It is common knowledge that the analyticity properties of the Feynman amplitudes can lead to remarkable results. This issue is rooted in the history of Quantum Field Theory (QFT) starting from the works of Kramers discussed at the Como conference in 1927, 
where, in particular, the connection between microcausality of local fields and analyticity of amplitudes has been pointed out \cite{Como}. In this way it is possible to obtain very important {\it dispersion relations} which have been exploited, for example, in the 
late 1950s and 1960s by Chew, Mandelstam, Regge and others in an attempt to develop a theory of strong interaction without using the concept of local fields. Although the results of these authors were remarkable, it became evident in 1970s and 1980s that 
QCD is definitely a preferable description of strong interaction. However, dispersion relations are still important for the field theorist today (for example in the analysis of Deep Inelastic Scattering). Recently, the analytic properties of amplitudes have been 
discussed in connection to modern techniques for the calculation of scattering amplitudes (see e.g. the textbook \cite{Elvang:2015rqa}).

In this article we will be particularly interested in the review of analyticity properties of Feynman amplitudes presented by Weinberg in his Cambridge books \cite{Weinberg:libro}. In particular, using Weinberg's notations and conventions, we will use the following 
claim \cite{Weinberg:libro}, section 10.2:

\

\noindent {\bf Claim:}\\
{\it Let us consider the momentum space amplitude $G$ given by
\begin{align}
 G(q_1,\ldots, q_n)=\int d^4x_1\cdots d^4 x_n e^{-i\sum_{j=1}^n x_j\cdot q_j}\ _0\langle T\{A_1(x_1)\ldots A_n(x_n)\} \rangle_0,
\end{align}
where $_0$ refers to the true vacuum $\Psi_0$ of the interacting theory and $A_j$ are Heisenberg operators with arbitrary spin, or local functions of the fields and their derivatives, so that bound states may be included.
If we group together a certain number $r$ of external lines (with $1 \leq r\leq n-1$) and
if we define the 4-momentum $q$ as
\[
q \equiv q_1+...+q_r=-q_{r+1}-...-q_n,
\]
then two results can be proved:\\
1. $G$ has a pole at $q^2=-m^2$, where $m$ is the mass of any 1-particle state $\Psi_{{\bf p}, \sigma}$ satisfying two conditions, namely\\
i) $\lambda_{1,\sigma} \equiv (\Psi_0, T\{A_1(x_1)...A_r(x_r)\}\Psi_{{\bf p},\sigma})\neq 0$,\\
ii) $\lambda_{2,\sigma} \equiv (\Psi_{{\bf p},\sigma}, T\{A_{r+1}(x_{r+1})...A_n(x_n)\} \Psi_0)\neq0$.\\
As usual, $\sigma$ represents the spin state (and any other internal quantum number) of the particle of mass $m$ and energy $p^0 \equiv \sqrt{{\bf p}^2+m^2}$.\\
2. The residue of $G$ at this pole is 
\be
G \longrightarrow \frac{-2i\sqrt{{\bf q}^2+m^2}}{ q^2+m^2-i \epsilon} (2 \pi)^7 \delta^4(q_1+...+q_n) \sum_\sigma M_{1,\sigma}(q_2,...,q_r) M_{2,\sigma}(q_{r+2},...,q_n)
\label{10.2.3}
\ee
where the $M$s are, by definition, given by
\be
(2\pi)^4 \delta^4(q_1+...+q_r-p) M_{1,\sigma}(q_2,...,q_r) \equiv \int d^4x_1...d^4x_r e^{-iq_1 \cdot x_1}...e^{-iq_r \cdot x_r} \lambda_{1,\sigma}
\label{10.2.4}
\ee
and
\be
(2\pi)^4 \delta^4(q_{r+1}+...+q_n+p) M_{2,\sigma}(q_{r+2},...,q_n) \equiv \int d^4x_{r+1}...d^4x_n e^{-iq_{r+1} \cdot x_{r+1}}...e^{-iq_n \cdot x_n} \lambda_{2,\sigma}.
\label{10.2.5}
\ee}
\hfill $\Box$\\

Our aim in this letter is to 1) show that the optical theorem, at all orders in perturbation theory, is a simple consequence of this general fact, so providing a new proof independent on cutting rules; 
2) discuss a new proof of the standard result concerning the effective action as generating functional of 1PI contributions. 

The optical theorem is a standard result of non-relativistic potential scattering theory \cite{Weinberg:2013qm}. It can be extended to relativistic QFT \cite{Weinberg:libro} where, in particular, it is typically presented in its generalized form which is valid 
order-by-order in perturbation theory (see e.g. \cite{Schwartz}):

\

\noindent {\bf Claim (Generalized optical theorem):}
\be
M(i \rightarrow f)-M^*(f \rightarrow i)=i \sum_X \int_\Delta \prod_{j \in  X} \frac{d^3p_j}{(2\pi)^3}\frac{1}{2E_j} (2\pi) \delta^4(p_i-p_X) M(i\rightarrow X)M^*(f \rightarrow X).
\label{GOT}
\ee

\hfill $\Box$\\

Some comments are in order:\\
1) $\Delta$ is the momentum resolution;\\
2) the theorem in its generalized form can be proved also at all orders of perturbation theory using cutting rules;\\
3) The special case $|i\rangle=|f\rangle$ corresponds to the well-known optical theorem of scattering theory. \\

It is our intention to prove the optical theorem at all orders of perturbation theory in the special case $| i\rangle=| f\rangle$. Interestingly, our proof will be particularly simple. %We will use Weinberg's notations and conventions.

In the final section we will analyze the second point mentioned above, namely we will present a new proof that the effective action is the generating functional of 1PI contributions.

%%%%%%%%%%%%%%%%%%%%%%%%%%%%%%%%
\section{A new proof of the optical theorem at all orders}

{\bf Claim:}\\
{\it Let $| A\rangle$ be some state such that $| i\rangle=| f\rangle=| A\rangle$, then it is possible to prove at all orders of perturbation theory that
\[
Im M(A \rightarrow A)\propto 2E_{CM}\sum_X \sigma (A \rightarrow X).
\]
}
{\bf Proof:}\\
We start from \eqref{10.2.3}
\be
G \longrightarrow \frac{-2i\sqrt{{\bf q}^2+m^2}}{ q^2+m^2-i \epsilon} (2 \pi)^7 \delta^4(q_1+...+q_n) \sum_\sigma M_{1,\sigma}(q_2,...,q_r) M_{2,\sigma}(q_{r+2},...,q_n)
\ee
and we apply this formula to the special case $r=2$, $n=4$. This is representative of the choice $| i\rangle=| f\rangle=| A\rangle=|$2-particle state$\rangle$ (but, as we will see, our claim is valid also with states that are not necessarily 
2-particle states). To proceed further, we use the well known Lehmann-Symanzik-Zimmermann (LSZ) reduction formula in the M factors to obtain formally the replacement $| \Psi_0\rangle \rightarrow |$2-particles$ \rangle$. Then, we have just to notice that \\
1) we have one factor $M$ relating the two particles in the initial state with the 1-particle intermediate  state (which is supposed to be a bound state); \\
2) we have another factor $M$ relating the intermediate single particle state with the final 2-particle state; \\
3) a factor $2E_{CM}=2\sqrt{{\bf q}^2+m^2}$; \\
4) we put the propagator $1/(q^2+m^2-i\epsilon)$ on-shell in harmony with \eqref{10.2.3} and, hence, the $i \epsilon$ term develops the imaginary part of the amplitude. \\
If we identify the cross section with the product of the $M$ factors we proved the claim, because when we take the square of the amplitude we are free to 
exchange the initial state with the intermediate one. In this way the process 2-particles$\rightarrow$intermediate (bound) state$\rightarrow$ 2-particles, becomes $| A \rightarrow $bound state of 2-particles$|^2$. Now we interpret the bound state as the 
{\it sum} over all possible states $X$: this is correct because each factor $M$ is 1/2 of the composite bridge and we find an infinite number of different particles (exchanged inside the bridge). In other words, the infinite number of contributions of perturbation 
theory correspond to all possible final states $X$. Since \eqref{10.2.3} is valid at all orders of perturbation theory, this completes the proof. \hfill $\Box$

{\bf Remark:} We have assumed that the intermediate state is a bound state in order to keep everything as simple as possible, but it is not a strictly necessary condition. From our claim one can extract the precise form of the optical theorem, e.g. as stated
in \cite{Weinberg:libro}, by carefully employing the relations between $G$ and $M$, and between $|M|^2$ and the total cross section $\sigma_i$ of the $in$ state $|i\rangle$. In a more intuitive way: the procedure of taking the modulus squared of (2-particles $\rightarrow$ intermediate state) corresponds to a ''cut in the bridge'' and, in this sense, the states exchanged to form the bound state are ''converted'' in asymptotic states (and the bound state ''disappears'').

%%%%%%%%%%%%%%%%%%%%%
\section{A new proof of 1-irreducibility}

Our proof is a modified version of the Zinn-Justin's proof discussed in \cite{Zinn-Justin:libro}.

\

\noindent{\bf Claim:}\\
{\it The effective action generates all the 1PI (and only the 1PI) connected graphs with arbitrary powers of $\phi$.}

\noindent{\bf Proof:}\\
Let us perturb the action:
\be
S_\epsilon[\phi]=S[\phi]+\frac{\epsilon}{2}\sum_{i,j}\phi_i \phi_j=S[\phi]+\frac{\epsilon}{2}(\sum_i \phi_i)^2.
\ee
If we imagine a discretized spacetime, we can use a simple index $i=1,...,N$ instead of the spacetime point coordinate $x$. With this $N$-point lattice our formulas will be more compact. We will also use the notation $J_i\equiv j(x_i)$ so there will be no 
difference between a (continuous) lower case source and an upper case (discretized) one. In this compact notation, integrals over continuous spacetime coordinates are replaced by sums over repeated discrete indexes. The quadratic part of the action can 
be written as a modification to the propagator:
\be
\frac{1}{2} \sum_{i,j} \phi_i ({\cal K}_{ij}+\epsilon M_{ij})\phi_j
\ee
where $M_{ij}=1, \ \forall i,j$. Hence, the free propagator $\Delta_\epsilon$ is now the inverse of ${\cal K}+\epsilon M$:
\be
(\Delta_\epsilon)_{ij}=\Delta_{ij}-\epsilon v_i v_j + {\cal O}(\epsilon^2)
\ee
where $v_i \equiv \sum_j \Delta_{ij}$.
This formula is interesting because it tells us that the perturbation (at first order) produces {\it disconnected} (i.e. factorized) contributions to the propagator. Indeed, the matrix $M_{ij}$ is not the identity and, hence, there are also off-diagonal contributions.
The two disconnected terms $v_i$ and $v_j$ arise from the splitting of $M_{ij}$ into a sum of a diagonal part plus an off-diagonal part. This disconnected contribution will ``propagate'' inside $W$, the generating functional of the connected correlation
functions. Since this disconnected term has two vertices, 
we can generate it with a factorized (i.e. disconnected) ``double derivative'' of the form 
\be
\frac{\delta W}{\delta J(x)} \frac{\delta W}{\delta J(y)}.
\label{disconnectedcont}
\ee
What happens if we perform a Legendre transformation? Formally we are ``moving to the phase space'' and, needless to say, a phase space is always even-dimensional. This means that a Legendre transformation builds a {\it duality} between $J$ and $\phi$: 
whatever will be the source $J$ we consider, we must introduce also a corresponding $\phi$ field. After the Legendre transform has been performed, $J$ and $\phi$ are linked together. This tells us immediately that the Legendre dual $\Gamma$ 
consists entirely of connected diagrams, because the disconnected contribution \eqref{disconnectedcont} is not duality invariant. Indeed, the derivative $\delta W/\delta J$ is mapped into $\phi$ and not into a dual derivative term.\hfill $\Box$

\section{Conclusions}

Despite we considered very well known results in QFT, we suggested a new fresh and modern perspective on these theorems. In particular, we provided a proof of the optical theorem at all orders in perturbation theory as a consequence of pole-ology, and also have emphasized the role of ``duality'' in our (proof of the) theorem of the effective action. Interestingly, duality arguments are characterizing a relevant part of modern theoretical physics, but they are also rooted in the history of Science (consider for example the well-known de Broglie's duality). These proofs are particularly suitable for degree students tackling QFT.

%%%%%%%%%%%%%%%%%%%%%%%%%%%%%%%%%%%%%%%%%%%%%%%%%%%%%%%%%%%%%%%%%%%%%%%%%%%%%%%%%%%%%%%%%%%%%%%%%%%%%%%%%%%%%%%%%%%%%
% ACKNOWLEDGEMENTS
%%%%%%%%%%%%%%%%%%%%%%%%%%%%%%%%%%%%%%%%%%%%%%%%%%%%%%%%%%%%%%%%%%%%%%%%%%%%%%%%%%%%%%%%%%%%%%%%%%%%%%%%%%%%%%%%%%%%%
%\vspace{0.5cm}

%%%%%%%%%%%%%%%%%%%%%%%%%%%%%%%%%%%%%%%%%%%%%%%%%%%%%%%%%%%%%%

%%%%%%%%%%%%%%%%%%%%%%%%%%%%%%%%%%%%%%%%%%%%%%%%%%%%%%%%%%%%%%%%%%%%%%%%%%%%%%%%%%%%%%%%%%%%%%%%%%%%%%%%%%%%%%%%%%%%%
% BIBLIOGRAPHY
%%%%%%%%%%%%%%%%%%%%%%%%%%%%%%%%%%%%%%%%%%%%%%%%%%%%%%%%%%%%%%%%%%%%%%%%%%%%%%%%%%%%%%%%%%%%%%%%%%%%%%%%%%%%%%%%%%%%%

\providecommand{\href}[2]{#2}\begingroup\raggedright\endgroup
\end{document}